\documentclass[reprint,amsmath,amssymb,aps,prb,floatfix]{revtex4-2}
\usepackage[utf8]{inputenc} 
\usepackage{graphicx}
\usepackage{dcolumn}
\usepackage{bm}
\usepackage{comment}
\usepackage{amsmath,amssymb,physics,bm,bbm}
\usepackage[safe]{tipa}
\usepackage[caption=false]{subfig}
\usepackage[hidelinks]{hyperref}
\usepackage{nameref}
\usepackage{orcidlink}
\usepackage{overpic}

\begin{document}
	
	\preprint{APS/123-QED}
    \title{Magnon-induced hybridization brightens excitons in ferromagnetic semiconductors}
	
	\author{Man-Yat Chu\orcidlink{0009-0003-7986-8709}}
	\email{mchu03@phas.ubc.ca}
	\affiliation{Quantum Matter Institute, University of British Columbia, Vancouver, British Columbia V6T 1Z4, Canada}
	
	\affiliation{Department of Physics and Astronomy, University of British Columbia, Vancouver, British Columbia, Canada, V6T 1Z1}
	
	\author{Mona Berciu}
\affiliation{Quantum Matter Institute, University of British Columbia, Vancouver, British Columbia V6T 1Z4, Canada}

\affiliation{Department of Physics and Astronomy, University of British Columbia, Vancouver, British Columbia, Canada, V6T 1Z1}
	
\date{\today}

\begin{abstract}
Excitons in magnetic semiconductors have energies and spin structures that are sensitive to the underlying magnetic order. We study this coupling in a minimal one-dimensional lattice model of a conduction electron and a valence hole moving in a ferromagnetic background of localized quantum spins, treating the electron, the hole, and the magnon(s) as explicitly resolved degrees of freedom. At zero temperature, spin conservation closes the relevant Hilbert subspaces at the one- or two-magnon level, so the model can be solved essentially exactly with a real-space Green's function method. Comparison against a frozen-spin approximation in which the local moments are replaced by their ordered values identifies which effects are due to emission and absorption of quantum magnons. A frozen background already lifts the spin degeneracy of the exciton and, if the two carriers couple with different strengths to the local moments, it mixes the singlet with the spin-zero triplet exciton. The emission and absorption of quantum magnons goes qualitatively further: it splits and shifts the exciton energies nonlinearly in the exchange coupling even when all frozen-spin effects vanish; for carriers with unequal hopping integrals, it hybridizes the singlet and spin-zero triplet excitons, thereby brightening the dark triplet; and it increases the exciton radius, which in turn enhances the magnon dressing by removing an on-site cancellation. These quantum effects are therefore strongest for extended, Wannier-like excitons, which is precisely the regime that is commonly modeled assuming a static (frozen-spin) magnetic order.
\end{abstract}

\maketitle

\section{Introduction}
\label{sec:Intro}

Magnetically ordered semiconductors and insulators support optical excitations whose properties depend on the surrounding spin order.  When an electron is promoted from the valence band to the conduction band, it can bind into an exciton with the resulting hole.  Because both carriers carry spins that interact with the local moments, the exciton's properties are affected by the magnetism in two distinct ways. First, the ordered moments act as a static field that shifts exciton energies, for instance lifting some of the triplet degeneracy but also hybridizing the $S_z^\mathrm{exc}=0$ triplet and singlet excitons.
Second,  the electron and hole can emit and absorb magnons, distorting the magnetic order in their vicinity. Such processes dress them into quasiparticles with renormalized properties, known as spin-polarons~\cite{ShastryMattis1981}; they also generate a magnon-mediated interaction that renormalizes the bare electron-hole attraction. Both of these affect the properties of the resulting excitons in ways that cannot be mimicked within a static magnetic scenario.

Exciton--magnon coupling has been studied for several decades.  Magnon sidebands of exciton lines were observed early on in the optical absorption of antiferromagnetic insulators such as MnF$_2$~\cite{SellGreeneWhite1967} and explained by the theory of magnon-assisted electric-dipole transitions~\cite{TanabeMoriyaSugano1965}. Moreover,  excitons and magnons were shown to form bound states whose interaction modifies these magnon-assisted optical transitions~\cite{FreemanHopfield1968}.  In ferromagnetic semiconductors such as the europium chalcogenides, the strong shift of the absorption edge upon magnetic ordering demonstrated how sensitive optical excitations are to the underlying spin order~\cite{MaugerGodart1986}.  More recently, optically active magnetic excitations have been observed in several van der Waals materials, see Ref.~\onlinecite{Brennan2024}. For example, the narrow, spin-correlated exciton in NiPS$_3$ is strongly tied to its antiferromagnetic background~\cite{Kang2020}.  Another example is CrSBr, which provides a particularly clear platform for the interaction between excitons and collective spin dynamics.  Its exciton energies respond strongly to the relative magnetization of neighbouring layers~\cite{Wilson2021}; coherent magnons modulate its exciton resonances~\cite{Bae2022}; and this response can be controlled by magnetic field~\cite{Diederich2023}, strain~\cite{Cenker2022}, and nonlinear magnon excitation~\cite{Diederich2025}.  The electronic structure of CrSBr is strongly anisotropic,  quasi-one-dimensional~\cite{Klein2023,Bianchi2023}, and recent measurements indicate the coexistence of small Frenkel-like excitons and of more extended, Wannier--Mott-like excitons~\cite{Smiertka2026}.  These materials motivate the question: how does an ordered, quantized magnetic background modify the internal structure of an exciton?

Available theoretical methods have focused on various aspects of this problem.  First-principles calculations retain realistic bands, orbital character, electron--hole binding, and optical selection rules, but these calculations are commonly carried out for a prescribed, static, quasi-classical magnetic order~\cite{Qian2023,Acharya2026}.  They therefore describe some of the dependence of the exciton properties on the magnetic order, but do not allow the emission and reabsorption of magnons by the exciton's electron and hole.  Within a model Hamiltonian approach, magnons can be kept as dynamical degrees of freedom while the exciton has been represented by a single composite quasiparticle.  This approach has been used to describe transport phenomena such as magnon--exciton drag~\cite{Iakovlev2026}.  However, projecting the electron--hole pair onto a rigid exciton envelope freezes their internal dynamics and does not allow, for instance, the magnons to modulate the exciton radius. Finally, we note that spin-mediated binding, formation, and recombination of Hubbard--Mott excitons have been studied in strongly-correlated single-orbital Hubbard models~\cite{Wrobel2002ExcitonsMottInsulators,Huang2023SpinMediatedMottExcitons,Lenarcic2013UltrafastRecombination}, where the exciton is a bound state between a doubly occupied and an empty site (a doublon--holon pair) interacting with the spin excitations of the same orbital. The doublon and the holon carry no spins, hence this problem is qualitatively different from that of interest to us, namely the effect of the magnetic order on excitons in regular (not strongly correlated) insulators.

The complementary microscopic problem of one or few electrons (holes) moving in an otherwise empty (full) band and coupled to a quantum magnetic background is also well established. Exact Green's function calculations have shown how the carrier emits and absorbs magnons to form a spin-polaron, and how magnon exchange can produce effective interactions between same-type carriers~\cite{ShastryMattis1981,BerciuSawatzky2009,Moller2012}. Correlations between carriers and magnons have also been studied in double-exchange ferromagnets~\cite{Kapetanakis2006ThreeBodyCorrelations}.  These calculations treat carriers and magnons as independent objects, but they do not describe exciton bound states because they only include one type of carrier. 

A methodological gap therefore remains for the treatment of exciton--magnon coupling as a few-body problem that explicitly resolves all three constituents: the electron, the hole, and the magnon(s).  As we show below, this gap matters especially when (i) the electron and hole have different effective masses, (ii) the exchange of magnon(s) between the electron and hole is significant, or (iii) the exciton extends over a few lattice sites.

Here we relax these limitations by studying a minimal one-dimensional (1D) lattice model, which describes a conduction electron, a valence hole, their Coulomb attraction, and local exchange interactions between both carriers and a chain of localized quantum spins with ferromagnetic {(FM)} order.  The FM background is the simplest setting in which the few-magnon sectors close exactly (see Sec.~\ref{sec:methods}). It is also physically motivated, because the individual layers of CrSBr are internally FM~\cite{Wilson2021} and their electronic structure is quasi-1D~\cite{Klein2023,Bianchi2023}, so a FM chain can be viewed as a minimal caricature of a single CrSBr layer. The 1D geometry suffices to expose the various consequences in a setting where this few-particles problem  (electron + hole + magnons) can be solved efficiently. We note that the solution we present below generalizes straightforwardly to describe more complex models  in higher dimensions.

We solve this model using a real-space Green's function method developed for few-particle lattice problems~\cite{Berciu2011,ChuBerciu2026}.  At zero temperature and in the $S_z^\mathrm{tot}=NS$ sector reached starting from a singlet electron+hole pair added to the FM in its ground-state, spin conservation restricts the accessible Hilbert space to the {0 and 1}-magnon sectors, so the hierarchy of equations of motion closes exactly at the level of the magnon number.  The real-space problem is solved numerically using cutoffs for the electron--hole and carrier--magnon distances, chosen sufficiently large that convergence is achieved to desired accuracy.  Throughout, we compare this full solution with a frozen-spin calculation in which the transverse spin operators are removed and the longitudinal local moments are replaced by their ordered value.  This comparison separates ordinary exchange-field effects from the effects due to emission and absorption of quantum magnons.

We first show that a frozen magnetic background already affects the spin of the exciton. In a non-magnetic background, excitons have either singlet or triplet character, and the three triplet components are degenerate. Longitudinal $S^z s^z$ exchanges with the frozen magnetic background lift the triplet degeneracy and also hybridize the singlet with the spin-zero triplet into two exciton states without a well-defined spin.

We then identify effects driven by the quantum nature of the magnons, illustrating points (i)--(iii) above.  In the symmetric case where the two carriers' exchanges with the magnetic background are equal, the static mixing mentioned above vanishes, however, magnon emission and absorption still produce a nonlinear dependence of the exciton's eigenenergies and their associated spin on the exchange coupling.  When the electron and hole hopping integrals have unequal magnitudes [point (i)], the two magnon-mediated virtual paths no longer cancel, generating an off-diagonal singlet--triplet coupling with no frozen-spin counterpart [point (ii)].  The resulting transfer of singlet-projected spectral weight to a triplet-dominated pole offers, within a local spin-conserving dipole model, a possible route to brightening the dark spin-zero triplet exciton, of direct relevance to recent experiments accessing optically suppressed excitons in CrSBr~\cite{Bork2026}.  Finally, these effects are controlled by the exciton size [point (iii)]: a finite electron--hole separation removes the on-site cancellation that suppresses magnon emission, therefore larger, more Wannier-like excitons are more strongly dressed by magnons.

The remainder of this paper is organized as follows.  Sec.~\ref{sec:model} introduces the lattice model;  Sec.~\ref{sec:methods} presents the real-space Green's function formalism;  Sec.~\ref{sec:results} first validates the method in known limits and then discusses the static and dynamical magnetic effects; and Sec.~\ref{sec:conclusion} summarizes the results.

\section{The Lattice Model}
\label{sec:model}

We study a 1D chain with $N\to \infty$ sites and lattice constant $a=1$, whose  Hamiltonian is:

\begin{equation}
 \mathcal{H} =  \hat{T}_c + \hat{T}_v + \hat{U}  + \mathcal{H}_{\rm{FM}} + \mathcal{H}_{\rm c-FM} + \mathcal{H}_{\rm v-FM}. \label{eq:Total_H}
\end{equation}
Here $\hat{T}_c=-t_c \sum_{j,\sigma}[c^{\dagger}_{j+1,\sigma}c_{j,\sigma}+ \text{H.c.} ] + \Delta \sum_{j,\sigma} c^\dagger_{j,\sigma} c_{j,\sigma} = \sum_{k\sigma} E_c(k) c^\dagger_{k\sigma} c_{k\sigma}$ and  $ \hat{T}_v=-t_v \sum_{j,\sigma}[ v^{\dagger}_{j+1,\sigma}v_{j,\sigma}+ \text{H.c.} ]= \sum_{k\sigma} E_v(k) v^\dagger_{k\sigma} v_{k\sigma} $ describe the conduction and valence bands, respectively (for simplicity, we ignore hybridization between them).  The operators  $c_{j,\sigma}^\dagger$ and $v_{j,\sigma}^\dagger$ create a conduction and a valence band electron, respectively, at site $j$ with spin $\sigma$. Their Fourier transforms $c^\dagger_{k\sigma} = \frac{1}{\sqrt{N}}\sum_j e^{ikR_j} c^\dagger_{j,\sigma}$ and $v^\dagger_{k\sigma} = \frac{1}{\sqrt{N}}\sum_j e^{ikR_j} v^\dagger_{j,\sigma}$ create electrons in the {conduction and valence} band, respectively, with momentum $k$ and spin $\sigma$. The two bands' dispersions are $E_c(k) = - 2t_c \cos k + \Delta$ and $E_v(k) = - 2 t_v \cos k$.  $\Delta$ is used to adjust the band gap between these bands. 

For an undoped  semiconductor described by only $\hat{T}_c+ \hat{T}_v$, the ground-state is $|\mathrm{gs} \rangle= \prod_{k\sigma} v^\dagger_{k\sigma}|0\rangle$ with energy $E_\mathrm{gs}=\sum_{k\sigma}E_{v}(k)$. The lowest-energy excitations are electron-hole pairs $c^\dagger_{K+k,\sigma} v_{k,\sigma'} |\mathrm{gs}\rangle$ with energy $E_c(K+k)-E_v(k)+E_\mathrm{gs}$. For any given pair momentum $K$, these particle-hole excitations form a continuum starting from the gap energy  $E^\mathrm{gap}(K)= \min_k \big[E_c(K+k)-E_v(k) \big]$ above the ground state. For this simple model,
\begin{equation}
    E^\mathrm{gap}(K) = \Delta - 2\sqrt{t_c^2 + t_v^2 - 2 t_c t_v\cos K}
    \label{egap}
\end{equation}
We always choose $t_c>0$. As expected, if $t_v <0$ this is a direct gap semiconductor, with a minimum gap $E^\mathrm{gap}(K=0)= \Delta - 2(t_c-t_v)$. On the other hand, if $t_v >0$, this is an indirect gap semiconductor, with a minimum gap $E^\mathrm{gap}(K=\pi)= \Delta - 2(t_c+t_v)$.

The term $\hat{U}$ describes the repulsion between electrons {that reside} in the two bands, which generates the electron-hole attraction. For simplicity, we take this interaction to be purely on-site:
\begin{equation*}
    \hat{U} = -U\sum_{j, \sigma, \sigma'}c^\dagger_{j,\sigma} c_{j,\sigma} v_{j,\sigma'} v^\dagger_{j,\sigma'} ,
\end{equation*}
and note that generalization to any longer-range expression is trivial within the real-space formalism discussed below~\cite{ChuBerciu2026}. {The} addition of $\hat{U}$ to $\hat{T}_c + \hat{T}_v$ modifies the non-interacting spectrum described above by pulling a bound electron-hole eigenstate --the exciton-- inside the gap, below the electron+hole continuum. For this simple model, the exciton energy is
\begin{equation}
    E^{\rm exc}(K) = \Delta - \sqrt{U^2 + 4(t_v^2 + t_c^2 - 2 t_c t_v\cos{K})},
    \label{eexc}
\end{equation}
see Appendix~\ref{App:1D_exciton} for derivation, and its binding energy is $E^{\rm bind}(K)=E^{\rm gap}(K)-E^{\rm exc}(K)$. Because of the simplicity of the model, the spin-part of the exciton wavefunction can be either a singlet or a triplet; in other words, this exciton state is 4-degenerate.

Next, consider $\mathcal{H}_\mathrm{FM}$, which describes {the} interactions between the localized moments ${\bf S}_j$ located at sites $j$. We take these to have spin $S$ and to be subject to FM Heisenberg exchange with their nearest-neighbors:
\begin{equation*}
    \mathcal{H}_{\mathrm{FM}} = -J_{\rm FM}\sum_j \mathbf{S}_j\cdot \mathbf{ S} _{j+1}
\end{equation*}
The ground-state of $\mathcal{H}_{\mathrm{FM}}$ is $|\mathrm{fm}\rangle= | +S, +S\dots  , +S\rangle$ and its lowest-energy excitations are single magnons with momentum $q$ and energy $\Omega(q) = 4S  J_\mathrm{FM}\sin^2 (q/2)$, described by $S^-_q|\mathrm{fm}\rangle$ where ${S}^-_q = \frac{1}{\sqrt{2SN}}\sum_j e^{i q R_j} {S}^-_j$. 
 
In the absence of interactions between the local moments and the valence+conduction band electrons, the undoped ground-state of this FM semiconductor is $|\mathrm{FM}\rangle= |\mathrm{gs}\rangle \otimes |\mathrm{fm}\rangle$. It is gapless to spin-excitations (magnons)  and has gapped charged excitations (singlet and triplet excitons, plus the continuum of particle-hole excitations), as described above. 

Finally, we add Heisenberg exchanges between the electrons in either band and the localized moment at the same site:
\begin{equation*}
    \mathcal{H}_{b-\rm FM} =  J_b \sum_j \mathbf{ S}_j\cdot \mathbf{s}_{b,j} ,
    \label{a-FM}
\end{equation*}
where $b \in \{c,v\}$ are the band indices, $J_b$ is the corresponding electron-magnon exchange, and $\mathbf{s}_{b,j}$ is the spin of the electrons from band $b$ at site $j$, for example $s^+_{b, j}=b^\dagger_{j,\uparrow}b_{j,\downarrow}$, etc. 

For the electron promoted to the conduction band, $J_c>0$ means an antiferromagnetic (AFM) coupling to the local moments.  Using the relation between hole operators and electron operators $v^\dagger_{j,\sigma} = (-1)^{\frac12-\sigma} h_{j,-\sigma}$~\cite{CombescotShiau2016, LandauLifshitzQM}, $J_v>0$ also means that the coupling between the valence band hole and the local moments is AFM. 

To summarize, this minimal model has six independent parameters: $t_c, t_v, U, J_\mathrm{FM}, J_c, J_v$ (note that $\Delta$ is a simple energy shift {that sets} the gap energy). We choose $t_c=1$ as the unit of energy, but this still leaves a 5-dimensional parameter space to investigate for interesting new physics. Here we focus on the region of this parameter space where $t_v<0$ (direct gap insulator), $U>0$, and $J_v>0$, $J_c>0$ implying AFM couplings between the conduction electron and valence hole with the FM background. Our goal in the following is to illustrate some of the more striking features that we have identified so far. 

\section{Formalism}
\label{sec:methods}

We solve this model by suitably generalizing a zero temperature, real-space Green's function method developed for few-particle lattice problems~\cite{Berciu2011}, and also used to study excitons in systems with multiple bands~\cite{ChuBerciu2026} as well as spin-polarons in FM systems~\cite{BerciuSawatzky2009} and their magnon-mediated interactions~\cite{Moller2012}. 

We are interested in {the effect of magnetic order on} the simple exciton spectrum described above, see Eq.~\eqref{eexc}, as we turn on the exchanges $J_v,J_c$ to the FM background. 

The simplest case is for the triplet exciton with $S_z^\mathrm{exc}=+1$. Because both the conduction electron's and the valence hole's spins are 'up', only the Ising $S_z s_{b,z}$ parts of the carrier-FM exchanges $\mathcal{H}_{b-\mathrm{FM}}$ have a non-vanishing but trivial action on this exciton eigenstate, leaving its wavefunction unchanged but shifting its energy by $\big(J_c+J_v\big)S/2$. (Recall that $h^\dagger_{j\uparrow}= v_{j,\downarrow}$). 

More interesting is what happens to the other exciton eigenstates. Consider next, the singlet exciton. In the absence of coupling to the FM background, its eigenstate is a  linear combination of basis states of the type:
\begin{equation}
    \ket{K,S_{0M};R}
\equiv \sum_j \frac{e^{iKR_j}}{\sqrt{2N}}\,
        \big(c^\dagger_{j\downarrow}\,v_{j+R,\downarrow}
            + c^\dagger_{j\uparrow}\,v_{j+R,\uparrow}\big)\,\ket{\mathrm{FM}}, \label{S}
\end{equation}
(recall that $h^\dagger_{j\downarrow}= -v_{j,\uparrow}$). Beside the energy shift from the diagonal parts of $\mathcal{H}_{b-\mathrm{FM}}$, their off-diagonal parts now generate new states of the type:
\begin{equation}
\ket{K,T^+_{1M};R,m}
\equiv \sum_j \frac{e^{iKR_j}}{\sqrt{2SN}}\,
        S^-_{j+m}\,c^\dagger_{j\uparrow}\,v_{j+R,\downarrow}\,\ket{\mathrm{FM}}. \label{T+}
\end{equation}
by raising the spin of the spin-down carrier and creating one magnon (1M) in the FM background besides a component of a  $T^+$ exciton with $S_z^\mathrm{exc}=+1$. Acting again with $\mathcal{H}_{b-\mathrm{FM}}$ on these additional basis states links back to the singlet basis states of Eq.~\eqref{S}, but also to the electron-hole triplet $T^0$ with $S_z^\mathrm{exc}=0$ basis states:
\begin{equation}
\ket{K,T^0_{0M};R}
\equiv \sum_j \frac{e^{iKR_j}}{\sqrt{2N}}\,
        \big(c^\dagger_{j\downarrow}\,v_{j+R,\downarrow}
            - c^\dagger_{j\uparrow}\,v_{j+R,\uparrow}\big)\,\ket{\mathrm{FM}}, \label{T0}
\end{equation}
The Hilbert subspace generated by the basis states of Eqs. (\ref{S}), (\ref{T+}) and (\ref{T0}), is closed under the action of the Hamiltonian because  conservation of the total spin forbids the admixture of the $S_z^\mathrm{exc} =\pm 1$ triplet excitons.

One option is to directly diagonalize $\mathcal{H}$ in this Hilbert subspace, to find the new exciton eigenfunctions: 
\begin{equation}
|K {\rm ;\lambda} \rangle = \sum_{\alpha, x_\alpha} \phi_{K, \lambda} (\alpha, x_\alpha)\ket{K,\alpha; x_\alpha}.
\label{eq:exciton_state} 
\end{equation}
with $\alpha \in \{S_{0M}, T^+_{1M}, T^0_{0M} \}$ and we use the shorthand notation $x_{S_{0M}} = R$, $x_{T^0_{0M}}=R$ and $x_{T^+_{1M}} = (R,m)$. In principle both $R$ and $m$ span all values from $-N/2$ to $N/2$, however we are primarily interested in bound complexes of these particles, where the probability of large separations is exponentially small.  As a result,  we  introduce two cutoffs $R_{\rm max}$ and $M$ and set $|\phi|^2 \to 0$ when $|R|>R_{\rm max}$ and when $\min\{|m|,|R-m|\}>M$. The values of $R_{\rm max}$ and $M$ are then increased until convergence is reached to the desired {precision}. This limits the dimension of the basis set and one could proceed with direct diagonalization of $\mathcal{H}$ within this variational subspace. 

Instead, we prefer to calculate electron+hole propagators within  this variational subspace and extract the exciton eigenenergies and wavefunctions from them~\cite{ChuBerciu2026}, avoiding the need to diagonalize increasingly large matrices when {studying} Wannier-like excitons. To this purpose, we define the real-space Green's functions
\begin{equation}
G_{\alpha\beta}(K,x_\beta,z)
=
\langle K,\alpha;0_\alpha|
\hat G(z)
|K,\beta;x_\beta\rangle .
\label{8}
\end{equation}
Here, $0_\alpha$ means that $R=0, m=0$, {\em i.e.} we project on the bra with on-site singlet/triplet electron-hole pair for $\alpha=S_{0M}, T^0_{0M}$ and the on-site electron, hole and magnon,  $R=m=0$  state for $\alpha=T^+_{1M}$  in Eq.~\eqref{T+}. Here \(\hat{G}(z) = (z - \mathcal {H})^{-1}\) is the resolvent of the Hamiltonian, $z=\omega+i\eta$ is the energy plus a broadening $\eta\rightarrow 0^+$ and we set $\hbar=1$. As already mentioned, we measure all energies in units of $t_c=1$ throughout, unless otherwise specified.

According to the Lehmann representation ~\cite{Lehmann1954,Mahan2000}, within this variational subspace:
\begin{equation}
G_{\alpha,\beta}(K, x_\beta, z) = \sum_{\lambda} \frac{\langle K, \alpha, 0_\alpha | K,\lambda\rangle\langle  K,\lambda | K, \beta, x_\beta \rangle}{z- E(K,\lambda)}
\label{9}
\end{equation}
where ${\cal H}|K,\lambda\rangle= E(K,\lambda)|K,\lambda\rangle$ are the eigenstates of ${\cal H}$ with a total momentum $K$. The eigenenergies are therefore signaled by poles of the propagators, in particular the bound states are the discrete poles lying below all the various continua, including the electron-hole continuum (see discussion below). However, because of the term $\langle K, \alpha, 0_\alpha | K,\lambda\rangle$ in the numerator, only eigenstates with a finite overlap with the chosen bra $\langle K, \alpha, 0_\alpha| $ are 'visible' in the $G_{\alpha,\beta}(K, x_\beta, z)$; we will call their corresponding exciton energies $E_\alpha(K,\lambda)$. Going through all $\alpha$ symmetries ensures that all discrete eigenstates are found (those with finite overlap with all the  $\langle K, \alpha, 0_\alpha| $ states appear in all the $E_\alpha(K,\lambda)$ spectra).

To extract their corresponding wavefunction, we use the value of the Green's function at the discrete pole of interest. So long as the separation between this discrete eigenstate and the next one is larger than $\eta$, Eq.~\eqref{9} approximates to:
\begin{align}
G_{\alpha,\beta}(K, x_\beta, z)&\big|_{\omega \rightarrow E_{\alpha}(K,\lambda)} \approx \notag \\ &\frac{\langle K, \alpha, 0_\alpha | K, \lambda\rangle\langle  K, \lambda | K, \beta, x_\beta \rangle}{z - E_{\alpha}(K, \lambda)}
\label{11}
\end{align}
and thus, see  Eq.~\eqref{eq:exciton_state}, $\phi^*_{K,\lambda}(\beta, x_\beta)=\langle  K, \lambda | K, \beta, x_\beta \rangle =C G_{\alpha,\beta}(K, x_\beta, E_\alpha(K,\lambda)) $
where the normalization constant $C$ is obtained from $\sum_{\beta,x_\beta}| \phi_{K,\lambda}(\beta,x_\beta)|^2 =1$.

These Green's functions are found by solving their equations of motion (EOM), obtained from the matrix elements of the identity $\hat{G}(z)(z-{\cal H}) = 1$ between various basis states of the variational subspace:
\begin{equation}
z G_{\alpha,\beta}(K, x_\beta, z) = \delta_{\alpha \beta} \delta_{0_\alpha,x_\beta} + \langle \alpha, K, 0_\alpha | \hat{G}(z) {\cal H}|  K, \beta, x_\beta \rangle
\label{13}
\end{equation}
As discussed above, ${\cal H}\ket{K, \beta, x_\beta}$ couples to other basis states, bringing {other} Green's functions {into} this EOM. For instance, hopping of either carrier generates coupling to states with the same symmetry $\beta$ but $R \to R\pm 1$ and/or $m\to m\pm1$, while the off-diagonal terms in $\mathcal{H}_{b-\mathrm{FM}}$ couples to Green's functions with  $\beta' \ne \beta$. Thus, the EOM form a coupled set of linear equations whose solution gives  $G_{\alpha,\beta}(K, x_\beta, z)$ for all  $\beta, x_\beta$ in the variational subspace.  The detailed calculation when $\alpha=S_{0M}$ is presented in Appendix~\ref{App:GF_EOM_S}; the other $\alpha$ are obtained from the same system of coupled equations,  except the inhomogeneous term $\delta_{\alpha \beta} \delta_{0_\alpha,x_\beta}$ is changed appropriately.

Finally, we also need to consider the subspace generated starting from the triplet exciton $T^-$ with $S_z^{\mathrm{exc}}=-1$, {\em i.e.} where the total spin is $S_z^\mathrm{tot}=NS-1$. In this case, both carriers can raise their spins,  hence up to two magnons can be created in the FM background. The EOM for the corresponding Green's functions are generated and solved similarly, see Appendix~\ref{App:GF_EOM_T} for full details.

In the following, we will plot the spectral weights associated with these Green's functions, for instance:
\begin{equation}
    A_{\alpha,\beta}(K,x_\beta,\omega) = -\frac{1}{\pi} \Im\{G_{\alpha,\beta}(K,x_\beta,z)\} .
\end{equation}
In these plots, a discrete state is signaled by a Lorentzian of width $\eta$ centered at the eigenstate energy, and the total area under the peak is the residue in Eq.~\eqref{11}, from which the wavefunction can be extracted.

\begin{figure}[t]
\centering
\includegraphics[width=0.95\linewidth]{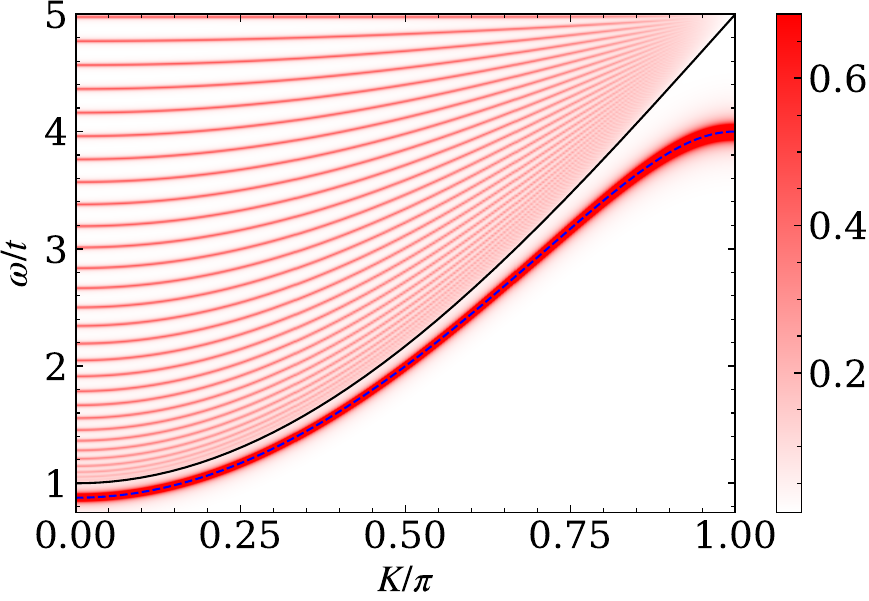} 
\caption{The heat map shows $A_{S_{0M},S_{0M}}(K,0_{S_{0M}},\omega)$, when  $t_c = -t_v = U = 1, \; \Delta =5 $ and $J_c = J_v = J_{FM} = 0$. The black line shows the expected lower-edge of the electron-hole continuum, Eq. (\ref{egap}), while the dashed blue line shows the expected exciton energy, Eq. (\ref{eexc}). Both are  in excellent agreement with the numerical results. }
\label{fig:ele_hole_exci}
\end{figure}

\section{Results}
\label{sec:results}

\subsection{Brief review of the known limits}

To set up the following discussion, and also as a way to validate our solution, we first briefly review known limits for this problem. 

\textit{Exciton not coupled to FM background:}  If we set $J_c = J_v = 0$, the carriers do not couple to the FM background and we should find the exciton with degenerate singlet and triplet configurations, whose energy is given in Eq.~\eqref{eexc}. This is verified in Fig.~\ref{fig:ele_hole_exci}, where we plot the heat map of the spectral weight obtained by projecting onto the singlet state $\alpha=S_{0M}$ (projecting on other symmetries produces the same spectrum). 

The black line shows the lower edge of the electron+hole continuum, Eq.~\eqref{egap}. The continuum that must lie above it is clearly visible, however it consists of many discrete peaks. This is a direct consequence of the cutoff $R_\mathrm{max}$, which constrains the maximum distance between electron and hole. This is equivalent to placing their relative motion inside an infinite potential well, resulting in discretization of the spectrum. Indeed, the average distance between peaks decreases with increasing $R_\mathrm{max}$, such that in the limit $R_\mathrm{max}\to \infty$ and/or for a sufficiently large $\eta$, the spectrum above the black line is indeed a smooth continuum.

Below this continuum there is one discrete state (broadened by $\eta$), showing the existence of an exciton whose energy is in perfect agreement with  Eq.~\eqref{eexc}, {illustrated} by the blue dashed line. Because we used a weak attraction $U=t_c=-t_v$, this exciton has a small binding energy and therefore a large exciton radius {$\xi$}. Its location starts to converge once {$R_\mathrm{max}> \xi$}, making this method especially efficient for studying smaller excitons.

\begin{figure}[t]
\centering
\begin{overpic}[width=\linewidth]{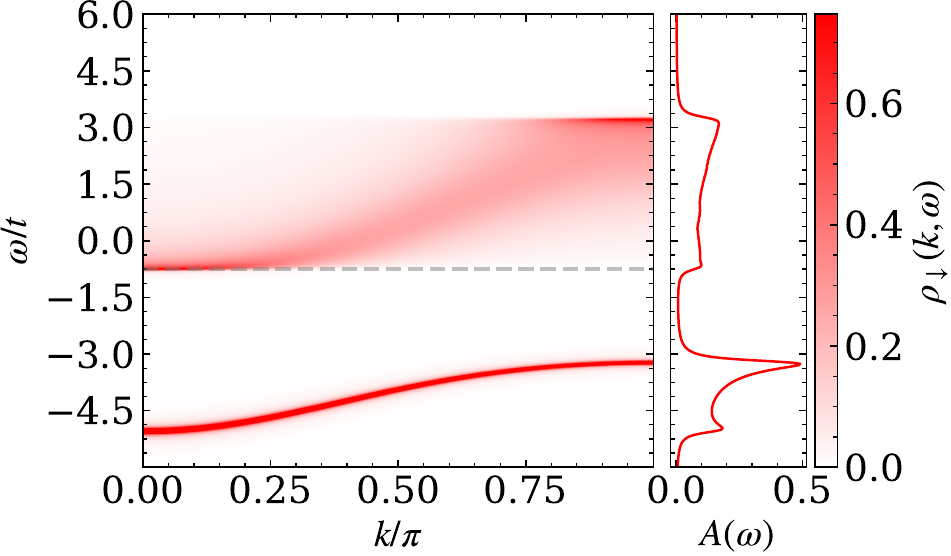}
    \put(60,12){\textbf{(a)}}
    \put(77,12){\textbf{(b)}}
\end{overpic}
\caption{(a) The heat map shows the spectral weight $\rho_\downarrow(k,\omega)$ for a single spin-down electron coupled to a FM background as reproduced from \cite{Moller2012}, with a discrete spin-polaron lying below the electron+magnon continuum; the predicted lower-edge of the latter is indicated by the dashed grey lines. Parameters are $J_{c} = 5$, $J_{FM} = 0.05$, $S = \frac12$ and $\eta = 0.02$. (b)~{The line graph} of our spectral weight $A(\omega) \equiv A_{T^+_{1M},T^+_{1M}}(K=0, 0_{T^+_{1M}},\omega)$, showing a lower continuum matching the spin-polaron bandwidth, and an upper continuum matching the electron+magnon bandwidth. See text for more details.}
\label{fig:Valid_sp}
\end{figure}

\textit{The spin-polaron limit} is reached when a spin-down single carrier with momentum $k$ (say, the electron in the conduction band) is coupled to the FM background. The off-diagonal part of the coupling can raise the carrier's spin and create a magnon with momentum $q$ in the FM background. These two objects can be unbound, giving rise to a continuum spanning $E_{c,\uparrow}(k-q) + \Omega(q) $; here and in the following, $E_{b,\sigma}(k)= E_{b}(k)+ \sigma J_b S $ is the energy of an electron with spin $\sigma$ in band $b=v,c$ shifted by the diagonal part of the coupling to the FM {background}. 

The carrier and the magnon can also bind, creating  a spin-polaron. For AFM  carrier-{local moment} coupling, the spin-polaron is the lowest-energy eigenstate, lying below the carrier+magnon continuum~\cite{ShastryMattis1981,BerciuSawatzky2009}. A typical spectrum for an electron spin-polaron~\cite{Moller2012} is shown in Fig.~\ref{fig:Valid_sp}(a), displaying the spin-polaron eigenstate well below the conduction electron+magnon continuum, whose expected lower edge is marked by the {grey} dashed line. 

Our formalism  can be used to verify this limit by setting $U = t_v = J_v = \Delta = 0$ so that the terms left in  Hamiltonian $\mathcal{H}$ describe the electron in the conduction band, the FM background, and their coupling.  However, by construction, our calculation also has a hole of infinite mass and zero energy, which acts as a scatterer: out of the total momentum $k$ of the system, the hole can carry any momentum $k'$, leaving $k+k'$ to the spin-polaron, or to the unbound electron and magnon in the continuum. This is confirmed by our calculation, see Fig.~\ref{fig:Valid_sp}(b), where we plot $A(\omega) \equiv A_{T^+_{1M},T^+_{1M}}(K=0, 0_{T^+_{1M}},\omega)$ for the same $t_c, J_\mathrm{FM}, J_c$ values as in panel (a). The lowest feature is a continuum matching the spin-polaron bandwidth, while the upper feature matches the electron+magnon continuum. Of course, one can repeat this for $U = t_c = J_c = \Delta = 0$ and check that the resulting hole spin-polaron appears at the expected energies.
 
Another way in which we checked this limit (not shown) is within the $S_z^\mathrm{tot}=NS-1$ subspace, where we start with both the electron and the hole with spins down. If we set $U=0$ then the electron and hole are not bound, and if we let $J_c, J_v\ne 0$, then both the electron and the hole will be dressed by their own magnon and turn into spin-polarons. We can calculate their individual dispersions $E_\mathrm{e-sp}(k)$ and $E_\mathrm{v-sp}(k)$ [similar to Fig.~\ref{fig:Valid_sp}(a)] and  verify that the lowest feature predicted by our calculation for any total momentum $K$ is the continuum spanned by $E_\mathrm{e-sp}(K+k)+E_\mathrm{h-sp}(k)$. The expected location of the higher continua are also known, for instance the highest is when all 4 objects become unbound, hence it lies at energies spanned by $E_{c,\uparrow}(K+k-q_1-q_2) - E_{v,\downarrow}(k) + \Omega(q_1)+\Omega(q_2)$.

\begin{figure}[t]
\centering
\begin{overpic}[width=0.472\linewidth]{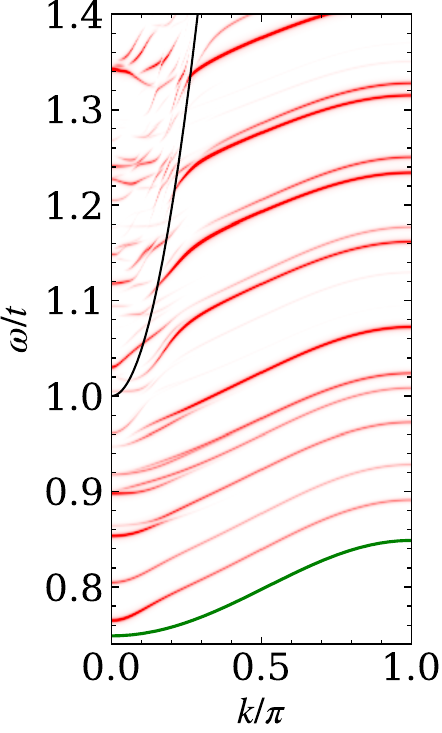}
    \put(45,15){\textbf{(a)}} 
\end{overpic}%
\begin{overpic}[width=0.518\linewidth]{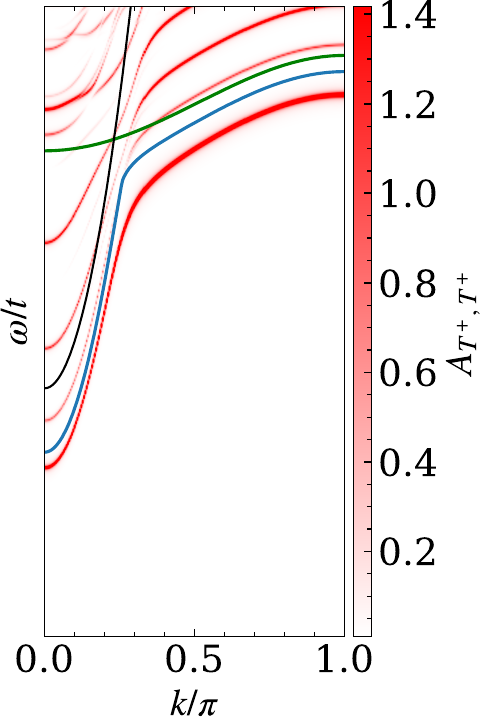}
    \put(36,15){\textbf{(b)}} 
\end{overpic}
\caption{ Heat maps of $A_{T^+_{1M},T^+_{1M}}(K, 0_{T^+_{1M}},\omega)$ for $U = 0.1$, $t_c =-t_v=  1$, $\Delta = 5, J_{FM} = 0.05$, $S = \frac12$, $\eta = 0.05$ when (a) $J_{c} = J_v = -0.5$, and (b) $J_{c} = J_v = 0.5$. The solid green line shows the lower edge of the exciton+magnon continuum, the black line is the lower edge of the electron+hole continuum, and the blue line is the lower edge of the electron spin-polaron+hole continuum. In (a) there is no discrete below the lowest continuum, hence here the low-energy states are unbound excitons and magnons. In (b) there is a discrete state below the lowest continuum, which is best thought of as a bound exciton between the undressed hole and an electron spin-polaron. See text for more details. }
\label{fig:Valid_cont}
\end{figure}

\subsection{How to interpret these spectra}

The various types of continua discussed above, whose bounds are known apriori, are useful not just to validate the code, but also to understand the nature of the lowest-energy eigenstates.

As an illustration, consider the subspace with $S_z^\mathrm{tot}=NS$, reached by adding either an electron-hole singlet or a $S_z^\mathrm{exc}=0$ triplet to the FM background. As discussed in Sec.~\ref{sec:model}, in this case there can be up to three objects in the system: the electron, the hole and a magnon. If all 3 are unbound, the resulting continuum spans $E_{c,\uparrow}(K+k-q)-E_{v,\uparrow}(k)+\Omega(q)$. 

There are various ways for 2 of the 3 objects to be bound, generating other possible continua with the 3rd unbound object; their ordering depends on the parameters. For instance, if the carrier-{local moment} couplings are FM then neither carrier will bind the magnon at low energies, however it is possible that the electron and hole bind into an exciton while the magnon remains free. The corresponding continuum could lie at low energies, starting at $ \min_q[E^\mathrm{exc}(K-q)+\Omega(q)]$. 

However, if either carrier-{local moment} exchange is AFM, then a low-energy spin-polaron can exist and one expects a low-energy continuum to form between this spin-polaron and the other carrier, {\em e.g.} the lower edge of the electron spin-polaron+hole starts at $\min_{k} [ E_{\rm e-sp}(K+k) - E_{v,\downarrow}(k)]$.

It is also possible that a magnon is not created and the electron and hole remain unbound, signalled by electron+hole continua spanning $E_{c\uparrow}(K+k)-E_{v,\uparrow}(k) $ and  $E_{c\downarrow}(K+k)-E_{v,\downarrow}(k) $ (these are identical if $J_c=J_v$). 

{The combination of the many parameters determines the lowest continuum.} Fig.~\ref{fig:Valid_cont} shows two possible cases. In panel (a), parameters are chosen such that the lowest continuum, whose lower edge is marked by the green line, corresponds to a $S_z^\mathrm{exc}=0$ exciton +  free magnon. There is no discrete state below this edge, meaning that for these parameters it is not possible to bind the electron+hole+magnon into one complex, instead the low {energy} states within this subspace consists of a free exciton and a free magnon. Note that changes in the eigenenergies inside a continuum indicate the edge of another continuum. For example, the black line marks the lower-edge of the electron+hole continuum.

In panel (b),  parameters are chosen such that the lowest continuum edge (thick blue line) corresponds to an electron spin-polaron+hole, while the green line marking the lower edge of the exciton+magnon continuum is at higher energy. The discrete state below these continua indicates a stable  electron+hole+magnon bound state, which is here best thought of as an 'exciton' that binds the undressed hole to the electron spin-polaron. 

Thus, knowledge of the location of the various continua is useful not just to validate the numerics, but also to indicate the best way to think of the nature of the  bound state (if any), as its character is inherited from  that of the lowest-energy continuum.

Before illustrating some of the interesting results, we remind the reader  that to understand the full spectrum, one needs to investigate all three possible Hilbert subspaces with $S_z^\mathrm{tot}=NS+1, NS$ and $ NS-1$.

\subsection{Splitting of the exciton spin states}
\label{sec:splitting}
\label{sec:nonlinear_magnon}

\begin{figure}[t]
\centering
\begin{overpic}[width=\linewidth]{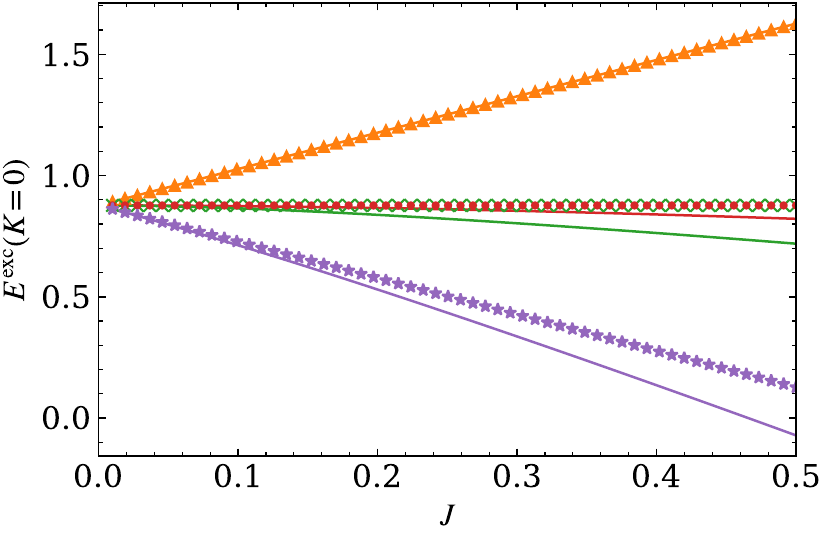}
    \put(17,15){(a)} 
\end{overpic}\\
\begin{overpic}[width=0.95\linewidth]{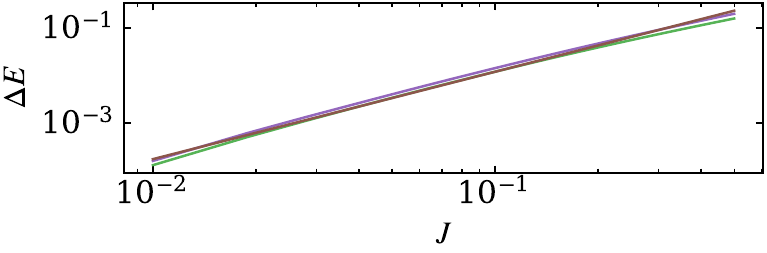}
    \put(20,20){(b)} 
\end{overpic}
\caption{(a) Energies of the lowest exciton with $K=0$ versus $J = J_c=J_v$, from the full calculation (solid lines) and the frozen-spin approximation (symbols): $T^+_{0M}$ (orange line and triangles), $T^-_{0M}$ (purple line and stars), $S_{0M}$ (red line and circles) and $T^0_{0M}$ (green line and crosses). The frozen-spin approximation is exact for the $T^+_{0M}$ exciton, however it captures only the linear part of the $T^-_{0M}$ shift, and predicts degenerate, $J$-independent $S_{0M}$ and $T^0_{0M}$ excitons (green crosses on top of red circles) because here $\Delta_J=0$. In the full calculation, virtual magnon exchange splits the $S_{0M}$ from the $T^0_{0M}$ exciton and shifts both of them, as well as the $T^-_{0M}$, down nonlinearly. (b) Log-log plot of the dynamical corrections $\Delta E$ between the full calculation and the frozen-spin approximation, versus $J$ for the $S_{0M}, T^0_{0M}$ and $T^-_{0M}$ excitons (same color scheme). The data fall on a straight line with slope 1.84, close to the value of 2 expected for second-order virtual magnon processes.
Parameters are $U = 1.0$, $-t_v=t_c = 1$, $\Delta = 5$, $J \in [0.01,0.5]$, $J_{FM} = 0.05$, $S = \frac32$ and $\eta = 0.01$. }
\label{fig:spin_plitting_non_lin}
\end{figure}

We now turn to the central question of this work: how does  the magnetic background modify the exciton states? Here we consider carriers with equal hopping magnitudes, $t_c=-t_v$, and equal exchange couplings, $J_c=J_v\equiv J$, because, as {discussed} below, this symmetric line is where the purely quantum effects of the magnons are exposed most cleanly. For each value of $J$, we locate the lowest discrete pole at $K=0$ in the spectral functions of the three $S_z^\mathrm{tot}$ sectors, projecting onto the $S_{0M}$, $T^0_{0M}$, $T^+_{0M}$ and $T^-_{0M}$ channels. The resulting eigenenergies are shown as the solid lines in Fig.~\ref{fig:spin_plitting_non_lin}(a).

The full calculation shows that the coupling to the magnetic background completely lifts the fourfold spin degeneracy of the exciton. The $T^+_{0M}$ exciton moves up linearly with $J$, while the $T^-_{0M}$ exciton moves down with a visible curvature. More remarkably, the $S_{0M}$ and $T^0_{0M}$ excitons split from each other even though the two carriers couple with equal strength $J$ to the background. Both are pushed below the bare exciton energy, with the $T^0_{0M}$ pushed considerably more so than the $S_{0M}$. These downward shifts are manifestly nonlinear in $J$.

To identify which of these features is of quantum origin, we compare against the frozen-spin approximation obtained by setting $S_j^z\rightarrow S$ and $S_j^\pm\rightarrow0$, {\em i.e.}, the transverse spin-flip processes are discarded while the longitudinal exchange acts on both carriers. This reduces the full theory to a mean-field description of the carriers in a static exchange field. In this limit, the $T^+_{0M}$ and $T^-_{0M}$ excitons acquire opposite energy shifts, linear in the exchange couplings, while in the subspace spanned by $\ket{K,S_{0M};R}$ and $\ket{K,T^0_{0M};R}$, the frozen-spin contribution is (see Appendix~\ref{App:GF_EOM_S}, restricted to this subspace):
\begin{equation}
H_{\mathrm{fr}}^{(S_{0M},T^0_{0M})}=
\begin{pmatrix}
0 & \Delta_J\\
\Delta_J & 0
\end{pmatrix},
\qquad
\Delta_J=\frac{(J_v-J_c)S}{2}.
\label{eq:static_ST0_matrix}
\end{equation}
The diagonal entries vanish because both spin-zero states contain equal weights of the two opposite carrier-spin configurations. The longitudinal exchange gives different energies to these two configurations, and this difference becomes the off-diagonal matrix element $\Delta_J$ in the singlet--triplet basis. The resulting frozen-spin eigenstates, with energies $E_{\pm}(K)=E^\mathrm{exc}(K)\pm |\Delta_J|$, are equal mixtures of $S_{0M}$ and $T^0_{0M}$, with the relative sign fixed by the sign of $J_v-J_c$. Thus, a frozen background can lift all spin degeneracies and mix the two spin-zero excitons {\em only if $J_c \ne J_v$}. On the symmetric line  $J_v=J_c=J$ the split $\Delta_J=0$, so the frozen-spin approximation predicts that the $S_{0M}$ and $T^0_{0M}$ excitons remain degenerate, at the $J$-independent bare energy of Eq.~\eqref{eexc} shown by the red circles and green crosses.

The comparison in Fig.~\ref{fig:spin_plitting_non_lin}(a) therefore cleanly separates the static from the dynamical effects. For the $T^+_{0M}$ exciton, the frozen-spin energies (orange triangles) fall exactly on top of the full calculation (orange line): in the $S_z^\mathrm{tot}=NS+1$ sector no magnon can be emitted, so here the frozen-spin description is exact. For the $T^-_{0M}$ exciton, the frozen-spin result (purple stars) captures the leading linear shift, but the full calculation (purple line) lies increasingly below it as $J$ grows: this exciton can emit up to two magnons, and their virtual emission and reabsorption lowers its energy further. For the spin-zero excitons the difference is qualitative: the frozen background predicts two degenerate, $J$-independent states (green crosses on top of the red circles), whereas the full calculation splits them and shifts them down. 

Fig.~\ref{fig:spin_plitting_non_lin}(b) quantifies these dynamical corrections, showing the log-log plot of the differences $\Delta E$ between the full results and the frozen-spin predictions  versus $J$. In all sectors this is a straight line with slope 1.84, close to the value 2, expected if the dominant process is second-order. This second order process originate as (i) a carrier flips its spin while emitting a magnon, (ii) the intermediate state propagates, and (iii) a second spin flip reabsorbs the magnon. The deviation from slope 2 is due to higher-order processes captured by the exact calculation, and to the finite fitting window.

The different magnitudes of the $S_{0M}$ and $T^0_{0M}$ corrections can be traced to the structure of the equations of motion, Eq.~\eqref{B4}. The zero-magnon Green's functions couple to the one-magnon ones through the combination of $J_c G_{T^+_{1M}}(z,K;R,0)$ (magnon emitted by the electron) and $J_v G_{T^+_{1M}}(z,K;R,R)$ (magnon emitted by the hole). For an on-site pair ($R=0$) these two amplitudes cancel exactly in the singlet channel when $J_c=J_v$, but add in the $T^0$ channel. Physically, this is because an on-site singlet has zero net spin and cannot exchange spin with the local moment. The singlet therefore acquires its magnon dressing only through the finite-$R$ components of its wavefunction, which explains why its shift in Fig.~\ref{fig:spin_plitting_non_lin}(a) is much smaller than that of the $T^0_{0M}$. This observation also anticipates the strong dependence of all these effects on the exciton radius, to which we return in Sec.~\ref{sec:Exc_rad}.

In the remainder of this work we therefore stay on the symmetric line $J_c=J_v$, where the frozen background produces neither splitting nor mixing of the spin-zero excitons, so that any such effect is a purely quantum-magnon effect.

\begin{figure}[t]
\centering
\includegraphics[width=0.95\linewidth]{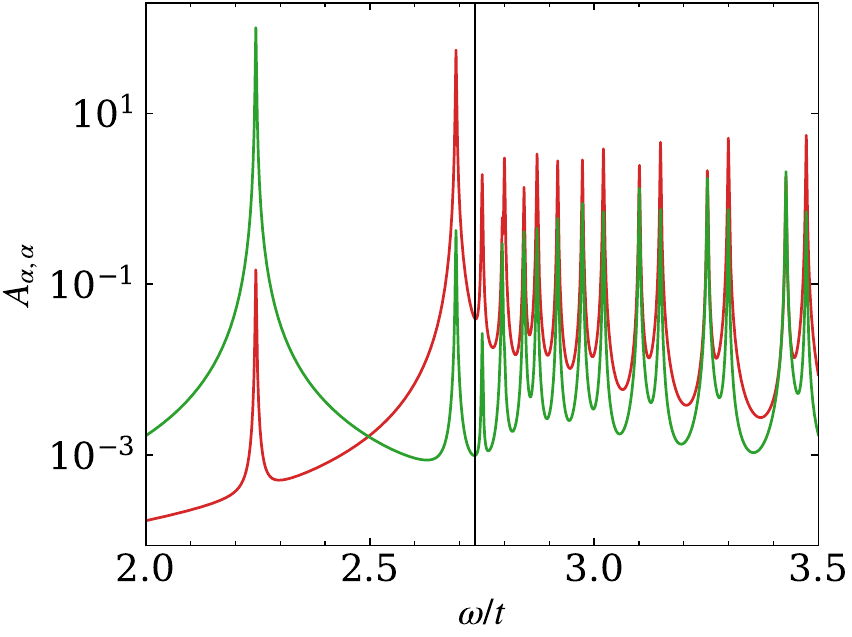} 
\caption{ Spectral functions $A_{\alpha,\alpha}(K=0,0_\alpha,\omega)$ for $\alpha=S_{0M}$ (red line) and $\alpha =T^0_{0M} $ (green line) when  $J_c=J_v$ but for unequal hopping magnitudes $t_c = 1.0$, $t_v = -1.4$. The edge of the lowest continuum {(electron spin polaron + hole in this case)} is shown by the vertical black line (the many sharp peaks above the continuum edge are continuum states discretized by the finite cutoff $R_\mathrm{max}$). The two discrete exciton states below the continuum carry weight in both projections, demonstrating the magnon-mediated singlet--triplet hybridization. For $t_c=-t_v$, electron--hole interchange symmetry  cancels this off-diagonal coupling in the exact calculation, while the frozen-spin approximation predicts no hybridization along $J_c=J_v$ for any hoppings. Other parameters are $U = 1.0$, $\Delta = 8.0$, $J_c = J_v = 1.5$, $J_{FM} = 0.05$, $S = \frac32$ and $\eta = 10^{-3}$. }
\label{fig:asymmetric_hopping}
\end{figure}

\subsection{Magnon-mediated singlet--triplet hybridization for unequal carrier masses}
\label{sec:asymmetric_hopping}

The previous subsection concerned exciton energies. We now examine the spin composition of the eigenstates, and ask whether the magnetic background can hybridize the $S_{0M}$ and $T^0_{0M}$ excitons on the symmetric line $J_c=J_v$. To achieve this, we need to break the electron--hole symmetry by taking unequal hopping magnitudes, $t_c\ne -t_v$.

\begin{figure}[t]
\centering
\begin{overpic}[width=\linewidth]{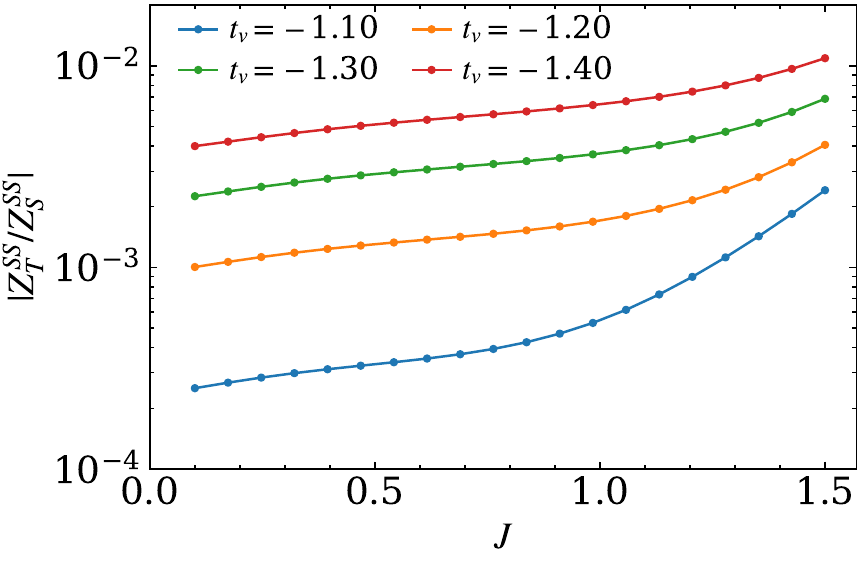}
\end{overpic}%
\caption{Ratio $|Z^{SS}_{T}/Z^{SS}_{S}|$ of the singlet-projected residues at the triplet-dominated and singlet-dominated poles, plotted against $J=J_c=J_v$ for several values of $|t_v|$, with $t_c = 1.0$. The hybridization grows monotonically both with $J$ which controls the magnon exchange, and with the hopping asymmetry. In the frozen-spin approximation this ratio is identically zero for any hopping integrals. Parameters are $U = 1.0$, $\Delta = 5$, $J_{FM} = 0.05$, $S = \frac32$ and $\eta = 10^{-4}$.}
\label{fig:residue_ratio_U_t}
\end{figure}

Figure~\ref{fig:asymmetric_hopping} shows the $S_{0M}$-projected ({red}) and $T^0_{0M}$-projected ({green}) spectral functions at $K=0$ for $t_c=1.0$, $t_v=-1.4$. The discrete poles below the continuum edge {(vertical black line; the lowest continuum is electron spin polaron + hole in this case)} carry weight in {\em both} projections: the $T^0_{0M}$-dominated exciton also appears as a small peak in the $S_{0M}$ projection at the same energy, and viceversa. The two spin-zero excitons are therefore hybridized by dynamic coupling to the magnetic background, even though $J_c=J_v$. The degree of hybridization is quantified in Fig.~\ref{fig:residue_ratio_U_t}, which plots the ratio $|Z^{SS}_{T}/Z^{SS}_{S}|$ of the singlet-projected residues at the triplet-dominated and singlet-dominated poles. This ratio grows monotonically with $J$ which drives the magnon exchange, and with the hopping asymmetry $|t_v|/t_c$.

None of this is captured by the frozen-spin approximation, where the only source of singlet--triplet mixing is $\Delta_J$ which vanishes for $J_c=J_v$. Because the hopping Hamiltonian is spin-independent, it acts identically on the $S_{0M}$ and $T^0_{0M}$ excitons, so no choice of $t_c, t_v$ can mix them. The hybridization found in the full calculation is instead an interference effect involving virtual magnons. There are two second-order paths connecting the two spin-zero excitons: in one, the electron emits the magnon and the hole reabsorbs it, while in the other, the roles of the two carriers are reversed. At $K=0$ and when $t_c=-t_v$ and $J_c=J_v$, the Hamiltonian has an electron--hole interchange symmetry. The two spin-zero excitons belong to different sectors of this symmetry and the two paths cancel exactly. Formally, the one-magnon Green's functions $G_{T^+_{1M}}(z,K;R,0)$ and $G_{T^+_{1M}}(z,K;R,R)$ entering the equation of motion, Eq.~\eqref{B4} of Appendix~\ref{App:GF_EOM_S}, are then identical, and while the two excitons still acquire different diagonal energies, they do not hybridize. Unequal hopping magnitudes, $t_c\ne -t_v$, break this cancellation. {After the magnon is emitted, there is an asymmetry between the configuration after the hopping of the electron and that of the hole; thus, the two Green's functions differ and an off-diagonal singlet--triplet coupling appears.} This coupling is a genuine dynamical effect, beyond linear order in $J$, and it vanishes continuously as $t_c\rightarrow -t_v$.

\subsection{Brightening of the dark triplet exciton}
\label{sec:bright_triplet}

The hybridization discussed above has a direct optical consequence, as it brightens the $T^0_{0M}$ exciton. To see this, consider local dipole coupling to light~\cite{haug_quantum_2005}:

\begin{equation}
    \mathcal{H}_I = \sum_{\sigma,i}E(t)( \bf{d}_{c,v}c^\dagger_{\sigma,i}v _{\sigma,i} +H.c.)
\end{equation}

The absorption coefficient is then given by \cite{dagotto_correlated_1994,Goodvin2011}:

\begin{equation}
    I(\omega) = -\frac{1}{\pi} \Im{\bra{\psi_0}\mathcal{H}_I\hat{G}(\omega)\mathcal{H}_I\ket{\psi_0}}.
    \label{eq:ab_coef}
\end{equation}
which is rewritten, using the Lehmann representation, as

\begin{equation}
    I(\omega) = \sum_n |\bra{\psi_0}\mathcal{H}_I\ket{\psi_n}|^2 \delta(\omega -E_n ).
\end{equation}

Here $\ket{\psi_0} = \ket{\rm FM}$, and because the dipole operator is local and spin-conserving, $\mathcal{H}_I\ket{\rm FM} \propto \big(c^\dagger_{0\downarrow}\,v_{0,\downarrow}+ c^\dagger_{0\uparrow}\,v_{0,\uparrow}\big)\, \ket{\mathrm{FM}} = \sqrt{2N} \ket{K=0, S_{0M},0}$, {\em i.e.}, light creates an on-site spin-singlet electron-hole pair. The absorption therefore measures the singlet-projected spectral function:

\begin{equation}
    I(\omega) \propto -\frac{2N}{\pi} \Im{G_{S_{0M},S_{0M}}(K=0,0,z)}.
\end{equation}

 Because of the magnon-mediated hybridization, the $T^0_{0M}$-dominated exciton carries a finite singlet-projected residue $Z^{SS}_{T}$, see Fig. \ref{fig:asymmetric_hopping}, which therefore appears in the absorption spectrum as a peak at the energy $E_{T^0_{0M}}$. The ratio plotted in Fig.~\ref{fig:residue_ratio_U_t} can thus be interpreted as the ratio of absorption intensities at the energies of the triplet-dominated and singlet-dominated excitons. Our results show that the dark triplet brightens progressively as the hopping asymmetry and/or $J$ increase. In contrast, in the frozen-spin description for $J_c=J_v$, the spin-zero triplet remains strictly dark for any choice of hoppings. This brightening is thus a distinct, optically accessible signature of the quantum magnon dynamics. Recent experiments demonstrating boson-enabled access to an optically suppressed exciton in CrSBr~\cite{Bork2026} suggest that such transfer of optical weight is experimentally relevant.

\subsection{Interplay between magnon dressing and the exciton radius}
\label{sec:Exc_rad}
\label{sec:U_tune_hybridisation}

\begin{figure}[t]
\centering
\begin{overpic}[width=\linewidth]{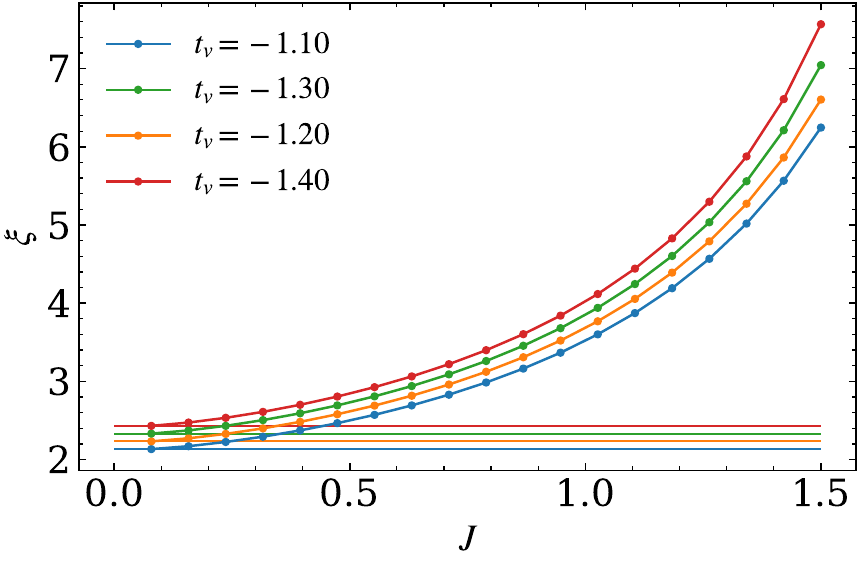}
\end{overpic}%
\caption{Exciton radius $\xi$ of the $S_{0M}$-dominated exciton versus $J=J_c=J_v$, for $t_c=1$ and $t_v=-1.1, -1.2, -1.3, -1.4$ (color scheme as in  Fig.~\ref{fig:residue_ratio_U_t}). The horizontal lines show the corresponding frozen-spin approximation predictions. In the full calculation the exciton radius grows monotonically with $J$, and faster for excitons with a larger bare radius (larger $|t_v|$), showing that magnon exchange between the carriers weakens their effective attraction. Parameters are $U = 1.0$, $\Delta = 5$, $J_{FM} = 0.05$, $S = \frac32$ and $\eta = 10^{-4}$.}
\label{fig:exciton_radius_U_t}
\end{figure}

Finally, we investigate how the presence of magnons affects the internal structure of the exciton, characterized by its radius $\xi$ (defined, for example, in Ref.~\onlinecite{ChuBerciu2026}). Figure~\ref{fig:exciton_radius_U_t} shows $\xi$ for the $S_{0M}$-dominated exciton as a function of $J=J_c=J_v$, for several hopping asymmetries. In the full calculation the radius grows monotonically with $J$, slowly at first and then steeply, roughly tripling over the range shown. The growth is faster for larger $|t_v|$, {\em i.e.}, for excitons that already start out larger at $J=0$ (the binding energy decreases with increasing $|t_v|$, see Eq.~\eqref{eexc}).

In the frozen-spin approximation, by contrast, the radius is $J$-independent (horizontal lines in Fig.~\ref{fig:exciton_radius_U_t}), because along $J_c=J_v$ the frozen background does not affect the spin-zero excitons at all. The increase of $\xi$ is therefore entirely due to the exchange of  magnons between the two carriers, which mediates an effective interaction. In this parameter regime, this effective interaction is repulsive and it loosens the exciton. A more detailed analysis of this magnon-mediated interaction is left for future work.

\begin{figure}[t]
\raggedleft
\begin{overpic}[width=0.96\linewidth]{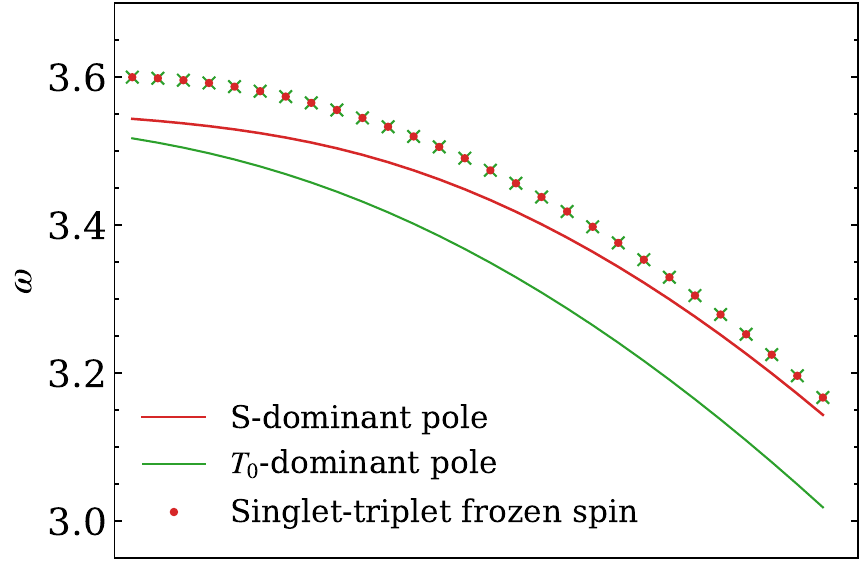}
\end{overpic}
\begin{overpic}[width=\linewidth]{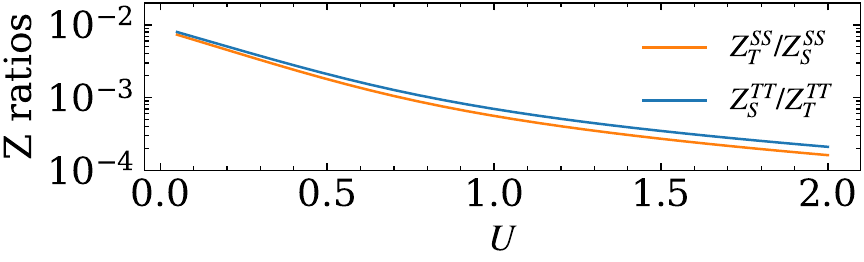}
\end{overpic}%
\caption{(a) Evolution with the Coulomb attraction $U$ of the $S_{0M}$-dominated exciton energy (red line) and of the $T^0_{0M}$-dominated exciton energy (green line), in the full calculation. The solid red dots trace the frozen-spin approximation prediction, which is a degenerate, unhybridized $S_{0M}$--$T^0_{0M}$ pair at all $U$. {(b) ratio of the residues $|Z^{SS}_{T}/Z^{SS}_{S}|$ and $|Z^{TT}_{S}/Z^{TT}_{T}|$ are plotted in blue and orange respectively.} Both ratios decrease with increasing $U$, because a larger $U$ binds the exciton more tightly and reduces its radius~\cite{ChuBerciu2026}, suppressing the magnon-mediated hybridization. Parameters are $U \in [0.01,2.00]$, $t_c = 1$, $-t_v = 1.2$, $\Delta = 5$, $J_c = J_v = 0.5$, $J_{FM} = 0.05$, $S = \frac32$ and $\eta = 10^{-4}$.}
\label{fig:U_scan_hybrid}
\end{figure}

Conversely, the exciton radius controls the strength of the magnon dressing. As discussed in Sec.~\ref{sec:splitting}, an on-site ($R=0$) singlet pair has zero net spin and its exchange with the local moments cancels when $J_v=J_c$.  Magnon-mediated effects on the singlet thus arise from the finite-$R$ components of its wavefunction, which are enhanced for larger excitons. Fig.~\ref{fig:U_scan_hybrid} tests this prediction by tuning the exciton radius through the Coulomb attraction $U$, at fixed $J$ and hopping asymmetry. Increasing $U$ binds the exciton more tightly and reduces $\xi$~\cite{ChuBerciu2026}. {Indeed, the singlet-projected residue ratio $|Z^{SS}_{T}/Z^{SS}_{S}|$ and the triplet-projected residue ratio $|Z^{TT}_{S}/Z^{TT}_{T}|$, plotted in Fig.~\ref{fig:U_scan_hybrid} (b), decreases monotonically as $U$ increases, confirming that smaller excitons hybridize less.} The frozen-spin approximation (solid circles) predicts that that the exciton remains an unhybridized, degenerate $S_{0M}$--$T^0_{0M}$ pair at all $U$.

Taken together, Figs.~\ref{fig:exciton_radius_U_t} and \ref{fig:U_scan_hybrid} reveal a positive feedback mechanism: magnon exchange expands the exciton, and a larger exciton couples more effectively to the magnons. The dynamical effects discussed in this work, due to the quantum nature of magnons,  are thus weaker for compact, Frenkel-like excitons and stronger for extended, Wannier-like excitons. The latter is precisely the regime in which excitons in magnetic semiconductors are commonly modeled assuming a frozen magnetic order.

\section{Conclusion}
\label{sec:conclusion}

In this work, we studied how coupling to a ferromagnetic background of localized quantum spins affects the spectrum and characteristics of excitons in a minimal 1D semiconductor model. We used a real-space Green's function method that resolves the electron, the hole, and the magnon(s) explicitly and is essentially exact at zero temperature. The method was validated against several known limits (the bare exciton, the spin-polaron, and the expected hierarchy of continua). Throughout, comparison with the frozen-spin approximation, in which the local moments are replaced by a static, fully ordered background, allowed us to identify which effects require the quantum dynamics of magnon emission and absorption.

The frozen background  lifts the fourfold spin degeneracy of the exciton: the $T^{\pm}$ triplets acquire opposite, linear energy shifts, and for $J_c\ne J_v$ the singlet hybridizes with the spin-zero triplet. The quantum magnons add three qualitatively new effects, all present even on the symmetric line $J_c=J_v$, where a frozen background leaves the spin-zero excitons degenerate. First,  magnon emission and absorption split the $S$ and $T^0$ excitons and shifts them apart, with an approximately quadratic dependence on the exchange coupling. Second, if the electron and hole hopping magnitudes differ, the two magnon-exchange paths no longer interfere destructively and the $S$ and $T^0$ excitons hybridize. Because optical absorption projects onto the singlet channel, this hybridization brightens the otherwise dark spin-zero triplet exciton even in the absence of spin-orbit coupling, phonons, mechanical strain or external magnetic fields. Third, the magnon exchange between the carriers weakens their effective attraction and increases the exciton radius, while a larger radius in turn enhances all magnon-mediated effects by removing an on-site cancellation in the singlet sector. This positive feedback implies that the quantum corrections are stronger for larger Wannier excitons, {\em i.e.}, precisely the regime where excitons in magnetic semiconductors are commonly described using the frozen background approximation.

Natural extensions of this work include studies of excitons in  multi-orbital models in  higher dimensions, with longer-range interactions and in the presence of magnetic fields, coupled not only to ferromagnetic but also to antiferromagnetic and canted backgrounds. 
We anticipate that such work will offer a more complete understanding of the possible effects of coupling to magnetic backgrounds on exciton spectra.

\section{Acknowledgments}
	
This project was undertaken thanks in part
to funding from the Max Planck-UBC-UTokyo Center for Quantum Materials and the Canada First Research Excellence Fund, Quantum Materials and Future Technologies Program, as well as the Natural Sciences and Engineering Research
Council of Canada (NSERC). We gratefully acknowledge the use of computing resources from the Stewart Blusson Quantum Matter Institute computing cluster LISA. Additionally, we acknowledge the valuable discussion with Prof. Adina Luican-Mayer. Lastly, we acknowledge that this research was done on the UBC Point Grey (Vancouver) campus, located on the traditional, ancestral, unceded territory of the $x^wm$\textipa{@}$\theta k^w$\textipa{@}$\overset{,}{y}$\textipa{@}$m$ (Musqueam) First Nation.

\newcommand{\Tp}{T^+}
\newcommand{\Tm}{T^-}
\newcommand{\Tz}{T^0}
\newcommand{\FM}{\mathrm{FM}}
\newcommand{\VB}{\mathrm{VB}}
\newcommand{\sep}{\mathrm{sep}}
\newcommand{\dbl}{\mathrm{dbl}}
\newcommand{\JFM}{J_{\mathrm{FM}}}
\newcommand{\Hspin}{H_{\mathrm{spin}}}
\newcommand{\Hsd}{H_{sd}}
\newcommand{\Heh}{H_{eh}}
\newcommand{\Emag}{E_{\mathrm{mag}}}
\newcommand{\supp}{\operatorname{supp}}
\newcommand{\sgn}{\operatorname{sgn}}

\appendix

\section{Binding energy of 1D exciton}
\label{App:1D_exciton}

For $J_c=J_v=0$, the magnetic and excitonic sectors decouple, and the problem reduces to an electron and a hole on a one-dimensional lattice with an on-site attraction.  Because the Hamiltonian is spin independent in this limit, the singlet and triplet excitons have identical spectra.  We therefore
suppress the spin label and write an eigenstate with total momentum $K$ as
\begin{equation}
    \ket{\Psi_K}
    =
    \sum_R \phi_K(R)\ket{K;R}.
\end{equation}
Using
\begin{equation}
    (\hat{T}_c + \hat{T}_v)\ket{K;R}
    =
    -B_+\ket{K;R-1}
    -B_-\ket{K;R+1},
\end{equation}
and
\begin{equation}
    {\hat{U}}\ket{K;R}
    =
    -U\delta_{R,0}\ket{K;R},
\end{equation}
the equation is then a recurrence relation
\begin{align}
    (E_K-\Delta )\phi_K(R)
    &=
    -B_+\phi_K(R-1) \notag\\
    &-B_-\phi_K(R+1)
    -U\delta_{R,0}\phi_K(0),
    \label{eq:pure_exciton_schrodinger}
\end{align}
where $B_\pm=t_c e^{\mp iK}-t_v$. Solving the recurrence relation, the exact exciton energy is
\begin{equation}
    E^{\mathrm{exc}}(K)
    = \Delta- 
    \sqrt{
        U^2+
        4\left(t_c^2+t_v^2-2t_ct_v\cos K\right)}.
    \label{eq:exact_exciton_dispersion}
\end{equation}
and the binding energy measured from the lower continuum edge is
 \begin{align}
      E^{\rm bind}(K) =  \sqrt{U^2 + 4(t_v^2 + t_c^2  - 2 t_c t_v\cos{K})}\nonumber \\- 2\sqrt{t_v^2 + t_c^2 - 2 t_c t_v\cos{K}}.
 \end{align}

\section{The equation of motion for singlet initial state in the $S_z^\mathrm{tot}=NS$ sector}
\label{App:GF_EOM_S} 

We use the short-hand notations:
\begin{align}
G_{S_{0M}}(z,K;R)     &\equiv \bra{K,S_{0M};0}\,\hat{G}(z)\,\ket{K,S_{0M};R},\\
G_{\Tz_{0M}}(z,K;R) &\equiv \bra{K,S_{0M};0}\,\hat{G}(z)\,\ket{K,\Tz_{0M};R},\\
G_{\Tp_{1M}}(z,K;R,m) &\equiv \bra{K,S_{0M};0}\,\hat{G}(z)\,\ket{K,\Tp_{1M};R,m},
\end{align}
and  $g \equiv {\sqrt{S}}/{2}$, $\Delta_J \equiv {(J_v-J_c)\,S}/{2}$.

The EOMs obtained from $(z-H)\hat{G}=\mathbbm{1}$ read:
\begin{align}
&\big[z + U\,\delta_{R,0} - \Delta \big]\,G_{S_{0M}}(z,K;R) = \notag\\
&\delta_{R,0}
 - B_+\,G_{S_{0M}}(z,K;R-1) - B_-\,G_{S_{0M}}(z,K;R+1) \notag\\
&\quad - \Delta_J\,G_{\Tz_{0M}}(z,K;R) \notag\\
&\quad - g\big[J_c\,G_{\Tp_{1M}}(z,K;R,0)
 - J_v\,G_{\Tp_{1M}}(z,K;R,R)\big].\label{B4}
\end{align}
\begin{align}
&\big[z + U\,\delta_{R,0}- \Delta \big]\,G_{\Tz_{0M}}(z,K;R) = \notag\\
&- B_+\,G_{\Tz_{0M}}(z,K;R-1) - B_-\,G_{\Tz_{0M}}(z,K;R+1) \notag\\
&\quad - \Delta_J\,G_{S_{0M}}(z,K;R) \notag\\
&\quad - g\big[J_c\,G_{\Tp_{1M}}(z,K;R,0)
 + J_v\,G_{\Tp_{1M}}(z,K;R,R)\big].\label{B5}
\end{align}
\begin{align}
&\big[z - 2\JFM S + E^{\Tp_{1M}}_{R,m}
      + U\,\delta_{R,0}- \Delta \big]\,G_{\Tp_{1M}}(z,K;R,m) = \notag\\
&\quad - \JFM S\,\big[
G_{\Tp_{1M}}(z,K;R,m{-}1) + G_{\Tp_{1M}}(z,K;R,m{+}1)\big] \notag\\
&\quad - t_c\big[
e^{-iK}G_{\Tp_{1M}}(z,K;R{-}1,m{-}1)\notag\\
 &\quad+ e^{iK}G_{\Tp_{1M}}(z,K;R{+}1,m{+}1)\big] \notag\\
&\quad + t_v\big[
G_{\Tp_{1M}}(z,K;R{-}1,m)
+ G_{\Tp_{1M}}(z,K;R{+}1,m)\big] \notag\\
&\quad - g\,J_c\,\delta_{m,0}\,\big[
G_{S_{0M}}(z,K;R)+G_{\Tz_{0M}}(z,K;R)\big] \notag\\
&\quad + g\,J_v\,\delta_{m,R}\,\big[
G_{S_{0M}}(z,K;R)-G_{\Tz_{0M}}(z,K;R)\big],\label{B6}
\end{align}
where
\begin{equation}
E^{\Tp_{1M}}_{R,m} = \frac{(J_c + J_v)\,S}{2}
              - \frac{J_c}{2}\,\delta_{m,0}
              - \frac{J_v}{2}\,\delta_{m,R}.
\end{equation}

Note that only the EOM for the $R=0$ singlet carries $\delta_{R,0}$; choosing a different initial state (bra) simply moves this inhomogeneous term to the corresponding EOM.

The closed system of Eq.~\eqref{B4}-\eqref{B6} is exact for general $S$ and arbitrary $J_c, J_v$. To solve it, we set  $G \to 0$ when $R>R_{\rm max}$ and when $\min\{|m|,|R-m|\}>M$. Physically, for bound states the asymptotic probability for particles to be  separated by a distance $d$ decays exponentially with $d$; truncating
the linear system at finite $R_{\max},M$ therefore incurs only an exponentially small error which can be minimized by increasing $R_{\max},M$. For a given $R$, the unknowns are the
two zero-magnon Green's functions $G_{S_{0M}}(z,K;R),
G_{\Tz_{0M}}(z,K;R)$, and the one-magnon $G_{\Tp_{1M}}(z,K;R,m)$, which we combine in the vector:
\begin{equation}
\mathbf g_R
\equiv
\begin{pmatrix}
G_{S_{0M}}(z,K;R)\\
G_{\Tz_{0M}}(z,K;R)\\
G_{\Tp_{1M}}(z,K;R,-M)\\
G_{\Tp_{1M}}(z,K;R,-M+1)\\
\vdots\\
G_{\Tp_{1M}}(z,K;R,R+M)
\end{pmatrix},
\end{equation}
so the full Green's functions vector is
\begin{equation}
\mathbf G
\equiv
\begin{pmatrix}
\mathbf g_{-R_{\max}}\\
\vdots\\
\mathbf g_{R_{\max}}
\end{pmatrix}.
\end{equation}

With this ordering, Eq.~\eqref{B4}-\eqref{B6} are rewritten as a sparse matrix equation 
\begin{equation}
\underline M(z,K)\,\mathbf G(z,K)=\mathbf e
\end{equation}
where $\underline M(z,K)$ is a matrix whose entries are read from the EOM, and the inhomogeneous term $\mathbf e$ has all entries of zero except for the $S_{0M}$ with $R=0$, where the entry is 1.

This matrix equation is solved by sparse inversion.

\section{The equation of motion for $T^-_{0M}$ initial state in the $S_z^\mathrm{tot}=NS-1$ sector}

\label{App:GF_EOM_T}

\subsection{Basis}
The zero-magnon, $T^-_{0M}$  basis states are
\begin{equation}
\ket{K,\Tm_{0M};R}
=\frac{1}{\sqrt{2N}}\sum_j e^{iKj}
 c^\dagger_{j\downarrow}v_{j+R,\uparrow}\ket{\FM} .
\label{eq:tminus-basis-Tm}
\end{equation}
The relevant one-magnon states in this subspace are:
\begin{align}
&\ket{K,S_{1M};R,m}\notag\\
&=\frac{1}{\sqrt{2SN}}\sum_j e^{iKj}S^-_{j+m}
\left(c^\dagger_{j\downarrow}v_{j+R,\downarrow}
+c^\dagger_{j\uparrow}v_{j+R,\uparrow}\right)\ket{\FM},
\label{eq:tminus-basis-S}\\
&\ket{K,\Tz_{1M};R,m}\notag\\
&=\frac{1}{\sqrt{2SN}}\sum_j e^{iKj}S^-_{j+m}
\left(c^\dagger_{j\downarrow}v_{j+R,\downarrow}
-c^\dagger_{j\uparrow}v_{j+R,\uparrow}\right)\ket{\FM} .
\label{eq:tminus-basis-T0}
\end{align}
We label two-magnon states by an ordered pair $(m,n)$ with $m\le n$. The
case $m<n$ is the \emph{separated} sector,
\begin{align}
&\ket{K,\Tp_{2M};R,m,n}_{\sep}\notag\\
&=\frac{1}{2S\sqrt{N}}\sum_j e^{iKj}
S^-_{j+m}S^-_{j+n}c^\dagger_{j\uparrow}v_{j+R,\downarrow}\ket{\FM},
\label{eq:tminus-basis-Tpsep}
\end{align}
while $m=n$, possible only for $S\ge 1$, is the same-site \emph{double} sector:
\begin{align}
&\ket{K,\Tp_{2M};R,m}_{\dbl}\notag\\
&=\frac{1}{\sqrt{4S(2S-1)N}}\sum_j e^{iKj}
\left(S^-_{j+m}\right)^2c^\dagger_{j\uparrow}v_{j+R,\downarrow}\ket{\FM} ,
\label{eq:tminus-basis-Tpdbl}
\end{align}

Throughout, every two-magnon label produced by the couplings below is reduced to this form: coordinates are sorted by smaller-index first, and coincident coordinates are routed to the double sector. We write $r\!\setminus\!x$ for the single coordinate left after deleting one magnon at site $x$ from the pair $r=(m,n)$, while $N_x(m,n)= \delta_{m,x}+\delta_{n,x}$.

We also define the short-hand coupling coefficient
\begin{equation}
C(a,b)=
\begin{cases}
\frac{2S-1}{2}, & a=b,\\
\frac{\sqrt{S}}{2}, & a\ne b .
\end{cases}
\end{equation}

We choose the initial state as $\bra{K,\Tm_{0M};0}$ and define
\begin{align}
G_{\Tm_{0M}}(z,K;R)          &=\bra{K,\Tm_{0M};0}\hat G(z)\ket{K,\Tm_{0M};R},\\
G_{S_{1M}}(z,K;R,m)            &=\bra{K,\Tm_{0M};0}\hat G(z)\ket{K,S_{1M};R,m},\\
G_{\Tz_{1M}}(z,K;R,m)        &=\bra{K,\Tm_{0M};0}\hat G(z)\ket{K,\Tz_{1M};R,m},\\
G_{\Tp_{2M}}(z,K;R,r)        &=\bra{K,\Tm_{0M};0}\hat G(z)\ket{K,\Tp_{2M};R,r},
\end{align}
covering both the separated ($r=(m,n)$, $m<n$) and double ($r=(m,m)$)
sectors. Below we suppress the $(z,K)$ arguments. As in
Appendix~\ref{App:GF_EOM_S}, the unit source $\delta_{R,0}$ sits on a single line, here at  $\Tm_{0M}$ with $R=0$. A different initial state requires moving the unit source to the corresponding line.

The EOMs in this sector are:
\begin{align}
&\left[z-\Delta-E^{\Tm_{0M}}_R+U\,\delta_{R,0}\right]G_{\Tm_{0M}}(R) \notag\\
&\quad +B_+G_{\Tm_{0M}}(R-1)+B_-G_{\Tm_{0M}}(R+1) \notag\\
&\quad -g J_cG_{S_{1M}}(R,0)+g J_cG_{\Tz_{1M}}(R,0) \notag\\
&\quad +g J_vG_{S_{1M}}(R,R)+g J_vG_{\Tz_{1M}}(R,R)
 =\delta_{R,0} .
\label{eq:tminus-EOMTm}
\end{align}
Because $E^{\Tm_{0M}}_R=-\Delta_+\equiv -\frac{(J_v+J_c)\,S}{2}$, the entry in the diagonal bracket is
$z-\Delta+\Delta_++U\,\delta_{R,0}$. Next, 
\begin{align}
&\left[z-\Delta-2\JFM S+U\,\delta_{R,0}\right]G_{S_{1M}}(R,m) \notag\\
&\quad +\JFM S\left[G_{S_{1M}}(R,m-1)+G_{S_{1M}}(R,m+1)\right] \notag\\
&\quad +t_c e^{-iK}G_{S_{1M}}(R-1,m-1)\notag\\
&\quad+t_c e^{+iK}G_{S_{1M}}(R+1,m+1) \notag\\
&\quad +t_v\left[G_{S_{1M}}(R-1,m)+G_{S_{1M}}(R+1,m)\right] \notag\\
&\quad -J^{S_{1M}}_{R,m}\,G_{\Tz_{1M}}(R,m) \notag\\
&\quad -g J_c\,\delta_{m,0}\,G_{\Tm_{0M}}(R)
       +g J_v\,\delta_{m,R}\,G_{\Tm_{0M}}(R) \notag\\
&\quad -C(m,0)J_c\,G_{\Tp_{2M}}(R,m,0)\notag\\
       &\quad +C(m,R)J_v\,G_{\Tp_{2M}}(R,m,R)
 =0 .
\label{eq:tminus-EOMS}
\end{align}
where
\begin{equation}
J^{S_{1M}}_{R,m}=\Delta_J+\frac{J_c}{2}\delta_{m,0}-\frac{J_v}{2}\delta_{m,R}.
\label{eq:ES}
\end{equation}

Similarly,
\begin{align}
&\left[z-\Delta-2\JFM S+U\,\delta_{R,0}\right]G_{\Tz_{1M}}(R,m) \notag\\
&\quad +\JFM s\left[G_{\Tz_{1M}}(R,m-1)+G_{\Tz_{1M}}(R,m+1)\right] \notag\\
&\quad +t_c e^{-iK}G_{\Tz_{1M}}(R-1,m-1)\notag\\
&\quad +t_c e^{+iK}G_{\Tz_{1M}}(R+1,m+1) \notag\\
&\quad +t_v\left[G_{\Tz_{1M}}(R-1,m)+G_{\Tz_{1M}}(R+1,m)\right] \notag\\
&\quad -J^{S_{1M}}_{R,m}\,G_{S_{1M}}(R,m) \notag\\
&\quad +g J_c\,\delta_{m,0}\,G_{\Tm_{0M}}(R)
       +g J_v\,\delta_{m,R}\,G_{\Tm_{0M}}(R) \notag\\
&\quad -C(m,0)J_c\,G_{\Tp_{2M}}(R,m,0) \notag\\
       &\quad-C(m,R)J_v\,G_{\Tp_{2M}}(R,m,R)
 =0 .
\label{eq:tminus-EOMT0}
\end{align}
\newpage
Consider now the two-magnon states. The magnon hopping term moves one spin lowering operator from site $a$ to a neighboring site $b=a\pm1$, affecting the normalization. We define a convenient  shorthand:
\begin{align}
\mathcal J_{a\to b}(r)&=\frac{\JFM}{2}
\sqrt{N_a(r)\,(2S-N_a(r)+1)}\;\notag \\
&\times \sqrt{(2S-N_b(r))\,(N_b(r)+1)} .
\label{eq:Jhop}
\end{align}
For a single magnon ($N_a=1,N_b=0$) this reduces to $\JFM S$, matching entries in the
one-magnon EOMs.

The two-magnon EOM reads
\begin{align}
&\left[z-\Delta-E^{(2)}_{\mathrm{mag}}(r)-E^{\Tp_{2M}}_{R,r}+U\,\delta_{R,0}\right]
 G_{\Tp_{2M}}(R,r) \notag\\
&\quad +t_c e^{-iK}G_{\Tp_{2M}}(R-1,r-{\bf1})
       +t_c e^{+iK}G_{\Tp_{2M}}(R+1,r+{\bf1}) \notag\\
&\quad +t_v\left[G_{\Tp_{2M}}(R-1,r)+G_{\Tp_{2M}}(R+1,r)\right] \notag\\
&\quad +\sum_{a\in\supp(r)}\sum_{b=a\pm1}
      \mathcal J_{a\to b}(r)\,G_{\Tp_{2M}}(R,r_{a \to b}) \notag\\
&\quad -\mathbf 1_{N_0(r)>0}\,C(0,r)\,J_c
      \left[G_{S_{1M}}(R,r\!\setminus\!0)+G_{\Tz_{1M}}(R,r\!\setminus\!0)\right] \notag\\
&\quad +\mathbf 1_{N_R(r)>0}\,C(R,r)\,J_v
      \left[G_{S_{1M}}(R,r\!\setminus\!R)-G_{\Tz_{1M}}(R,r\!\setminus\!R)\right]\notag\\
      &=0 
\label{eq:tminus-EOMTp}
\end{align}
where
\begin{equation}
E^{\Tp_{2M}}_{R,r}=\Delta_+-\frac{J_c}{2}N_0(r)-\frac{J_v}{2}N_R(r) .
\label{eq:ETpUnified}
\end{equation}
and
\begin{equation}
E_{\mathrm{mag}}^{(2)}(r) = \begin{cases}
    4\JFM S & \text{if } m \le n-2 \\
    4\JFM S-\JFM  & \text{if }  m = n-1 \\
    4\JFM S & \text{if }  m=n
\end{cases}
\end{equation}

Also, $r \pm {\bf 1}= (m\pm 1, n\pm 1)$, while $r_{a \to b}$ is to replace the magnon at $a$ by a magnon at $b$. Finally, $\mathbf 1_{N_x(r)>0}$ returns a 1 as long as $N_x(r)>0$, which is derived from either the electron or the hole absorbs a magnon from the $S^+ s^-$ term. 

After suitable truncations and grouping, these equations combine into a very sparse linear system which is solved similarly to Appendix~\ref{App:GF_EOM_S}.

\bibliography{exciton_magnon_verified}

\end{document}